# Evanescent-Wave Bonding Between Optical Waveguides

**M. L. Povinelli,[1,2,*] Marko Lončar,[3,*] Mihai Ibanescu,[1] Elizabeth J. Smythe,[3]**

**Steven G. Johnson,[4] Federico Capasso,[3] J. D. Joannopoulos[1]**

[1]Department of Physics and the Center for Materials Science and Engineering, Massachusetts

Institute of Technology, 77 Massachusetts Avenue,

Cambridge, Massachusetts 02139

[2]Edward L. Ginzton Laboratory, Stanford University, Stanford, California 94305

[3]Division of Engineering and Applied Sciences, Harvard University, 9 Oxford Street,

Cambridge, Massachusetts 02138

[4]Department of Mathematics, Massachusetts Institute of Technology,

77 Massachusetts Avenue, Cambridge, Massachusetts 02139

**\*Both authors contributed equally to this paper.**

Forces arising from overlap between the guided waves of parallel, microphotonic waveguides are calculated. Both attractive and repulsive forces, determined by the choice of relative input phase, are found. Using realistic parameters for a silicon-on-insulator material system, we estimate that the forces are large enough to cause observable displacements. Our results illustrate the potential for a broader class of optically-tunable microphotonic devices and microstructured artificial materials.

*OCIS codes:* **230.7370, 230.7380, 200.4880, 250.5300**

Electronic bonding, resulting from overlap between electron wavefunctions, is a ubiquitous phenomenon that gives rise to material structure. It is intriguing to consider an analogous phenomenon of "optical bonding," where overlapping modes of linear optical waveguides result in the attractive or repulsive force between the waveguides.

It is known that intense, "external" laser illumination can create forces between dielectric or metallic spheres, due to interactions between the spheres' induced dipole moments (1-5). The nature of the force (attractive or repulsive) depends on the relative phase of the dipole moments (3, 6, 7). Our work identifies forces arising from "internal" illumination by light traveling in coupled waveguides. Unlike in previous work, we show that the sign of the force can be changed from attractive to repulsive *at fixed frequency* simply by tuning the relative phase of the system's inputs. Already, research has shown that evanescent fields can be used to trap (8), manipulate (9), and propel (10) nanoparticles. Our work extends the study of optical forces to interacting, evanescent waveguide modes, suggesting that coupling forces could provide a mechanism for optical positioning and control within integrated optical devices.

We consider two parallel, silicon strip waveguides (refractive index $n = 3.45$) separated by a distance $d$, as shown in Figure 1(a). Each waveguide has a square cross section of dimensions $a \times a$. We will characterize the modes of the two-waveguide system as even or odd with respect to the mirror planes at $y$=0 and $z$=0, according to the symmetry of the *vector* field. Vector symmetry implies that individual electric field components parallel to a mirror plane are symmetric (anti-symmetric), and components perpendicular to a mirror plane are anti-symmetric (symmetric) in the case of even (odd) modes (11).

The dispersion relation for y-odd modes (odd with respect to $y$=0 plane) of the two-waveguide system is shown in Figure 1(b). Modes lying outside the light cone (shaded yellow) are guided. For infinite separation of the waveguides ($d/a = \infty$), no coupling occurs. Modes of the two, isolated waveguides are degenerate, and the dispersion relation is that of a single strip waveguide, shown as a solid, black line. As the waveguides come closer, the degeneracy is broken and two distinct eigenmodes appear. Data for a separation of $d/a = 1.0$ are shown by a dashed red line, and data for $d/a = 0.1$ are shown by a dotted blue line. Modes tending to lie above the isolated waveguide mode have $z$-odd symmetry ($E_y$ is anti-symmetric with respect to $z=0$) and modes lying below the isolated waveguide mode have $z$-even symmetry ($E_y$ is symmetric).

Assume that energy $U = N\hbar\omega$ is coupled into an eigenmode (frequency $\omega$ and wave vector $\vec{k}$) of the system of two waveguides separated by a distance $\xi$. An adiabatic change in separation $\Delta\xi$ will shift the eigenmode frequency by $\Delta\omega$ ($\vec{k}$ is conserved due to preservation of translational invariance) and result in the mechanical force

$$F = -\frac{dU}{d\xi}\bigg|_{\vec{k}} = -\frac{d(N\hbar\omega)}{d\xi}\bigg|_{\vec{k}} = -N\hbar\frac{d\omega}{d\xi}\bigg|_{\vec{k}} = -\frac{1}{\omega}\frac{d\omega}{d\xi}\bigg|_{\vec{k}} U \qquad (1)$$

on either waveguide (12). Negative values here correspond to attractive forces. We used Eq. 1 to calculate the optically-induced force for a dimensionless frequency of $va/c = a/\lambda = 0.2$. Results are shown as symbols in Figure 1(c). The force was also calculated using the Maxwell stress tensor (13, 6), using the full numerical eigenmode solution of the two-waveguide system (14), and is shown by solid lines. Excellent agreement was obtained. The dimensionless force per unit length $(F/L)(ac/P)$ is plotted as a function of dimensionless waveguide separation $d/a$ on the left and bottom axes; $c$ is the speed of light in vacuum and $P = v_g U/L$ is the total power transmitted through the coupled waveguides ($v_g$ is the group velocity). Numerical values of force per unit length can be obtained by substituting typical experimental parameter values. For $va/c = 0.2$, operation at $\lambda = 1.55\mu$m requires that $a = 310$nm. The right and top axes show the dependence of the force per unit length on separation for an input power $P = 100$mW.

For separations $d/a$ greater than about 0.3, the anti-symmetric mode gives a repulsive force, while the symmetric mode gives an attractive force. The force decays exponentially with increasing separation, as expected from a tight-binding picture (15, 16): the two waveguides couple through the overlap of their evanescent modal tails. Intuitively, the lower-frequency, symmetric mode is attractive since its frequency decreases as the waveguides come together. Using the Maxwell stress tensor formalism, it is easy to show that modal contributions add incoherently to the spatially-averaged force. The force can thus be tuned from attractive to repulsive at fixed frequency by controlling the relative phase of the waveguide inputs.

For smaller separations, the tight-binding picture is no longer accurate. For the anti-symmetric mode, the force is *nonmonotonic:* repulsive for large separations and attractive for small separations. The zero-separation limit values were calculated to be -1.49 for the symmetric ($y$-odd, $z$-even) mode and -21.54 for the antisymmetric ($y$-odd, $z$-odd) mode. The strong attractive force associated with the antisymmetric mode is due to the enhancement of $E_z$ inside the slit due to the boundary conditions at the interface. $\vec{E}$-field enhancement was previously discussed in (17) for modes with $y$-even symmetry. For larger separations, we found that the $y$-even, $z$-odd mode is attractive and the $y$-even, $z$-even mode is weakly repulsive. In the zero-separation limit, the $y$-even, $z$-odd mode is strongly attractive due to field enhancement: $(F/L)(ac/P)$=-19.55. The $y$-even, $z$-even mode is weakly attractive: $(F/L)(ac/P)$=-0.21.

We next estimate the deflection ($w$) that a waveguide experiences due to an attractive, optically-induced force ($y$-odd, $z$-even mode). Figure 2(a) shows a potential implementation of our two-waveguide system in silicon on insulator material. Two Si waveguides, each of cross-sectional dimensions $a \times h$, rest on a SiO$_2$ substrate with a free-standing section of length $L$. Each waveguide can be described as a double-clamped beam loaded with force per unit length $q$ (18) that can be approximated with a linear function for small separations (Figure 1(c)):

$$EI \frac{d^4 w}{dx^4} = -q(w) \approx -q_d + 2\alpha\omega, \qquad (2)$$

where $I = ha^3/12$ is the moment of inertia of the waveguide and $E = 169$GPa is the Young modulus of Si. We assume that the change in phase velocity with distance along the bent waveguide is adiabatic and causes no reflection, valid assumptions for ($L >> w_{max}$). Reflection from the ends of the suspended section is also neglected. In Figure 2(b) we plot the displacement as a function of position along the waveguides for P=100mW ($\lambda$=1.55μm). Displacement of $w_{max} \approx 20$nm is obtained at the center of suspended section of each waveguide when $L$=30μm , $h$=$a$=310nm and $d$=46.5nm. Figure 2(c) shows the dependence of $w_{max}$ on $L$ and $P$; longer suspended regions deflect more for the same input power. The required power can be significantly reduced by reducing the waveguide core size (e.g. P=50mW when $h$=$a$=264nm) as well as making tall and narrow waveguides ($h > a$). This increases the fraction of energy in the air region between the waveguides, increasing the strength of the interaction, and also decreases the moment of inertia. The opto-mechanical response can also be improved by exploiting the mechanical resonance (19) of the suspended beams. Exciting the structure with a laser beam pulsed at the resonant frequency (MHz range) will significantly increase the beam deflection for a given input power.

An intriguing possibility for further increasing the magnitude of the force is slow-light enhancement, since the force increases as $1/v_g$ for fixed input power (20). We note that the electrostatic force due to trapped or induced charges in Si waveguides is estimated to be at least an order of magnitude smaller than the optically-induced force. The Casimir-Lifshitz force is even smaller. Optical evanescent-wave bonding forces are not restricted to the specific geometry and material system studied above. We have calculated that significant forces also arise, for example, between coupled silica microspheres (12). More generally, the use of optical, guided-wave signals to reposition the constituent parts of microphotonic devices suggests a new class of artificial, microstructured materials in which the internal mechanical configuration and resultant optical properties are coupled to incoming light signals.

The authors thank M. Brenner, D. Iannuzzi, M. Troccoli, J. Munday, C. Luo, J. T. Shen, S. Fan, M. Lipson, and S. Götzinger for useful discussions. This work was supported in part by the MRSEC program of the NSF under Award No. DMR-9400334, DoD/ONR MURI grant N00014-01-1-0803 and DARPA grant HR0011-04-1-0032.

# FIGURE CAPTIONS

**Figure 1.** (a) High-index ( $n = 3.45$ ) linear waveguides with square cross-section $a \times a$ separated by distance $d$. (b) Dispersion relation for varying waveguide separations for modes with $y$-odd vector symmetry. Insets show E$_y$ of waveguide modes (d/a=1.0) for modes lying above and below the infinite-separation limit, indicated by the black, solid line. (c) Normalized force per unit length as a function of separation at fixed frequency $\nu a / c = a / \lambda = 0.2$. Left and bottom axes are in dimensionless units. Right and top axes are in physical units with $P$=100mW, $\lambda$=1.55μm, and $a$=310nm. Solid lines are calculated using the stress tensor method; symbols are calculated using Eq. 1.

**Figure 2.** (a) Schematic of suspended section of two coupled waveguides. (b) Displacement of each waveguide as a function of position along the waveguide. *L=30μm, h=a, d/a=0.15, a/λ=0.2, λ=1.55μm* and input power *P=100mW*. (c) Displacement at the center of the suspended section (nm) as a function of power and length of suspended section. Region shown in white is where waveguides are in contact.

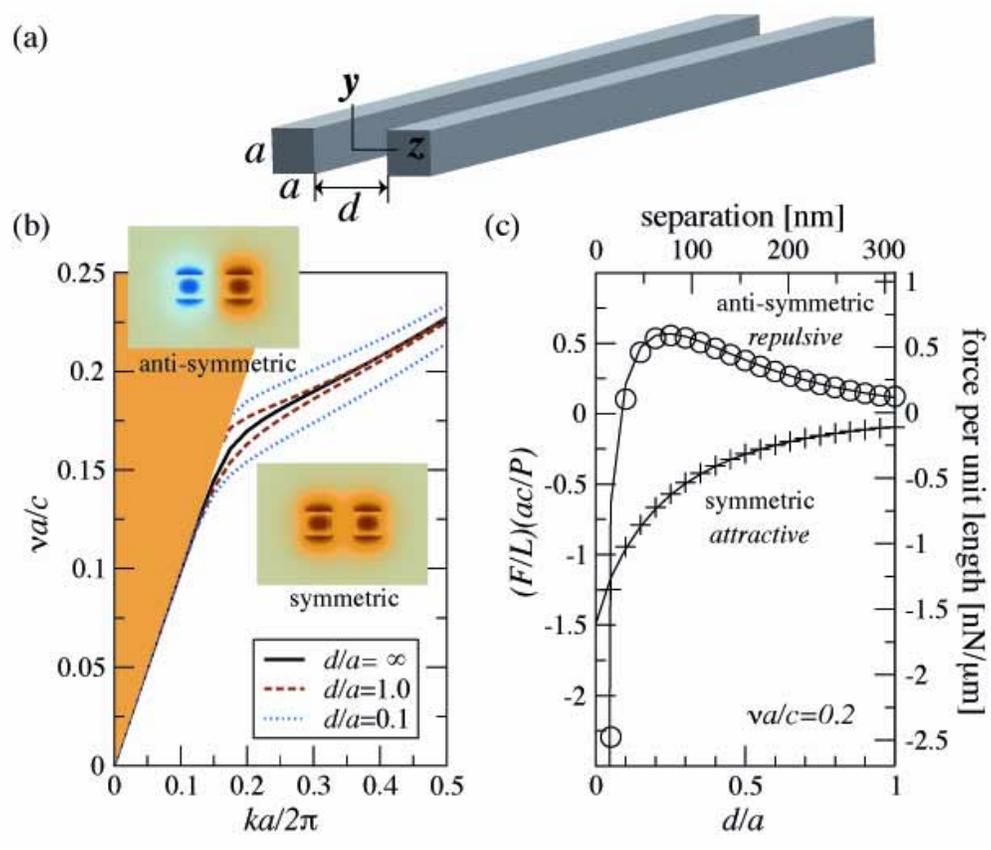

Figure 1

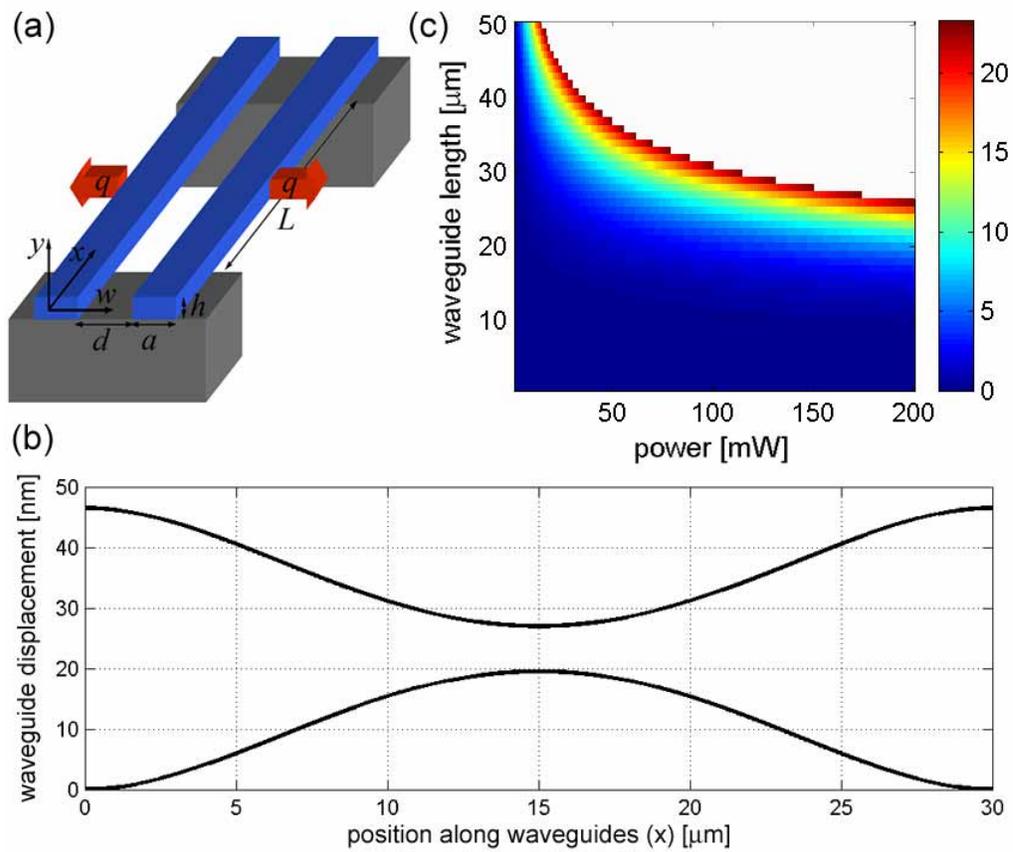

Figure 2